\documentclass[12pt]{article}
\usepackage{epsfig,axodraw}
\input epsf.sty
\input axodraw.sty
\newcommand{\la}[1]{\label{#1}}
 
\newcommand{\be}{\begin{equation}}
\newcommand{\ee}{\end{equation}}
\newcommand{\ba}{\begin{eqnarray}}
\newcommand{\ea}{\end{eqnarray}}
\newcommand{\bi}{\begin{itemize}}
\newcommand{\ei}{\end{itemize}}

\newcommand{\tr}{{\rm Tr\,}}
\newcommand{\ii}{{\rm i}}
\newcommand{\trq}{{\rm \hat{T}r\,}}
\newcommand{\ex}{{\rm e}}

\newcommand{\nn}{\nonumber}

\newcommand{\bfx}{{\bf x}}
\newcommand{\bfy}{{\bf y}}

\newcommand{\htm}{\hat{T}}
\newcommand{\op}{{\cal O}}

\newcommand{\RR}{{\rm I\kern -.2em  R}} 
\newcommand{\eq}{Eq.~}
\newcommand{\eqs}{Eqs.~}
\newcommand{\fig}{Fig.~}

\def\lsi{\raise0.3ex\hbox{$<$\kern-0.75em\raise-1.1ex\hbox{$\sim$}}}
\def\gsi{\raise0.3ex\hbox{$>$\kern-0.75em\raise-1.1ex\hbox{$\sim$}}}

\begin{document}
 
\begin{titlepage}
\begin{flushright}
MIT-CTP-3253
\end{flushright}
\begin{centering}
\vfill
 
{\bf NON-PERTURBATIVE FORMULATION OF THE\\ STATIC COLOR OCTET POTENTIAL}

\vspace{0.8cm}
 
Owe Philipsen

\vspace{0.3cm}
{\em 
Center for Theoretical Physics,\\ Massachusetts Institute of Technology,\\
Cambridge, MA 02139-4307, USA\\}

\vspace*{0.7cm}
 
\begin{abstract}
By dressing Polyakov lines with appropriate functionals of the gauge fields,
we construct observables describing a fundamental representation
static quark-antiquark pair in the singlet, adjoint and average channels of
SU(N) pure gauge theory.
Each of the potentials represents a gauge invariant eigenvalue of
the Hamiltonian.
Numerical simulations are performed for SU(2) in 2+1 dimensions.
The adjoint channel is found to be repulsive at small and confining
at large separations, suggesting the existence of a metastable 
$(N^2-1)$-plet bound state.
For small distances and temperatures above the deconfinement transition,
the leading order perturbative prediction for the ratio of singlet and adjoint
potentials is reproduced by the lattice data.
\end{abstract}
\end{centering}
 
\noindent
\vfill
\noindent
 

\vfill

\end{titlepage}
 

The potential of two static color 
charges in the fundamental representation separated by a distance $r$
is a quantity of fundamental importance for the study of the confining 
force in SU(N) Yang-Mills theory and QCD, as well as the phenomenology of
heavy quark systems.
A quark-antiquark pair at the same point
can be in either a singlet or an adjoint state,
acccording to the irreducible representations of the tensor product
\be
\bar{N}\otimes N=1 + (N^2-1).
\ee
In perturbation theory, one fixes a gauge and a separate description
for the two channels is possible also for small separations between the
charges. To leading order one finds the relation ($\op(g^2)$ contributions cancel)
\cite{bro,nad}
\be
V_1(r)=-(N^2-1)V_{N^2-1}(r) + \op(g^4).
\la{pert}
\ee
Because of the growing interaction strength, perturbation theory can
be trusted only at short distances, while non-perturbative
methods such as lattice simulations are required in the regime 
of linear confinement.

On the other hand, in lattice gauge theory one is
interested in non-perturbatively gauge invariant quantities
only\footnote{
Attempts were made to address this relation by lattice simulations in a
fixed Landau gauge \cite{att}. However, besides problems with uniqueness, 
lattice Landau gauge violates positivity
and it is not clear whether the results are associated with
physical quantum mechanical states.
}. It has so far not been possible to construct such operators 
for the two channels separately.
At zero temperature, the standard lattice operator to compute the 
static potential is the Wilson loop projecting 
onto the singlet channel. Good agreement between an integration 
of perturbatively truncated renomalization group equations 
and continuum extrapolated lattice results have been reported for separations
up to $\approx 0.3$fm \cite{som}.
At finite temperature, the only known gauge invariant operator
related to the potential is the Polyakov loop, which in turn yields
a weighted average of the singlet and adjoint channels \cite{ls,nad}. 
Thus, at asymptotically low temperatures, there are separate gauge invariant
operators for the singlet and average potential.
Together with the demonstrated 
validity of perturbation theory at short distances, this suggests 
that the octet channel exists also non-perturbatively.
 
Recently it has been shown that, at the expense of working with non-local 
functionals, it is possible to arrive at a gauge invariant 
and non-perturbative description of two-point functions for charge 
carriers, whose exponential decay is governed
by eigenvalues of the Hamiltonian \cite{op1}. This has been used 
to define and compute a parton mass for the gluon, which can be 
related to level splittings of static mesons \cite{op2}. 

In the present letter, we apply this formalism to
Polyakov loop correlators in order to establish the non-perturbative
existence and gauge invariance of the static potential 
in the singlet, adjoint and average channels.
After briefly summarizing 
the symmetries of a quark-antiquark pair, non-local eigenfunctions of the 
spatial lattice Laplace
operator are used to construct a dressed Wilson line, which is gauge invariant
up to a global residual symmetry of the Laplacian. 
Two such dressed Wilson lines separated by $r$ represent
an alternative operator to extract the zero temperature singlet potential, 
producing identical results to
those obtained from the Wilson loop.
In the finite temperature case, the dressed Polyakov loop allows for separately 
gauge invariant expressions for the average and the singlet potentials,
and the adjoint potential may be computed from the appropriate difference
between those operators.
The transfer matrix formalism is used to show the gauge
invariance of the potentials, and a numerical test in 2+1 d SU(2)
Yang-Mills theory demonstrates the feasibility of
simulating the operators in question.

We consider SU(N) Yang-Mills theory in $d+1$ dimensions 
with Wilson action on a $L^d\cdot N_t$ 
lattice. For a given Euclidean time, we are interested in the state with 
a static quark at $\bfy$ and its charge conjugate
at $\bfx$, $\bar{\psi}_\alpha(\bfx) \psi_\beta(\bfy)$.
The singlet and adjoint states of the pair are 
obtained by means of the projection operators \cite{bro,nad}
\ba
P_1&=&\frac{1}{N^2}\; 1\otimes 1 -\frac{2}{N}\;\bar{T}^a\otimes T^a,\nn\\
P_{ad}&=&\frac{N^2-1}{N^2}\; 1\otimes 1 + \frac{2}{N}\; \bar{T}^a\otimes T^a,
\ea
where $\bar{T}^a=-T^{a*}$ and
$
P_{1,ad}^2=P_{1,ad},\quad P_1P_{ad}=P_{ad}P_1=0,\quad P_{ad}+P_1=1\otimes 1.
$
Applying these operators to 
the quark-antiquark pair, one finds the tensor decomposition
\ba 
\left (P_1\bar{\psi}(\bfx)\psi(\bfy)\right)_{\alpha\beta}&=&\frac{1}{N}\delta_{\alpha\beta}
\bar{\psi}_\gamma(\bfx) \psi_\gamma(\bfy)\nn\\
\left (P_{ad}\bar{\psi}(\bfx)\psi(\bfy)\right)_{\alpha\beta}&=&
\bar{\psi}_\alpha(\bfx) \psi_\beta(\bfy)-\frac{1}{N}\delta_{\alpha\beta}
\bar{\psi}_\gamma(\bfx) \psi_\gamma(\bfy)\nn\\
&=&-2 \bar{T}^a_{\alpha\beta}\bar{\psi}_\gamma(\bfx)T^a_{\gamma\delta}\psi_\delta(\bfy).
\la{pair}
\ea
The projection operators act on the space of representation matrices, but 
are space-time independent. Therefore, under local gauge transformations,
the expressions \eq (\ref{pair}) behave as singlet and
$(N^2-1)$-plet for $\bfx=\bfy$ only. 

The propagation of a static quark in Euclidean time is described 
by the (untraced) temporal Wilson line \cite{po}, 
\be
\psi_\alpha(\bfx,t_0+t)=L(\bfx)_{\alpha\beta}\psi_\beta(\bfx,t_0),
\qquad L(\bfx)=\prod_{t'=t_0}^{t-1}U_0(\bfx,t'),
\la{line}
\ee
while that of a quark-antiquark pair is given by 
the matrix correlation function $C_{\bar{q}q}(r,t)$ of two 
Wilson lines separated by $r=|\bfx-\bfy|$.
According to the tensor decomposition
this can be written as a
sum of singlet and adjoint channels, 
\be
C_{\bar{q}q}(r,t)=\langle L^\dag(\bfx)L(\bfy)\rangle 
=\ex^{-t V_1(r)}P_1+\ex^{-t V_{ad}(r)}P_{ad}.
\ee
Applying the projection operators one solves for the 
respective channels
to find
\ba
\ex^{-t V_1(r)}&=&\tr(P_1C_{\bar{q}q})/\tr P_1
=\frac{1}{N}\langle \tr L^\dag(\bfx)L(\bfy)\rangle,\nn\\
\ex^{-t V_{ad}(r)}&=&\tr(P_{ad}C_{\bar{q}q})/\tr(P_{ad})\nn\\
&=&\frac{1}{N^2-1}\langle \tr L^\dag(\bfx) \tr L(\bfy)\rangle
-\frac{1}{N(N^2-1)}\langle \tr L^\dag(\bfx) L(\bfy)\rangle.
\label{potdef}
\ea
Again, these expressions are only gauge invariant for $\bfx=\bfy$.
This lack of gauge invariance is
a deficiency of the operators: so far they 
do not reflect the fact that between the $q\bar{q}$ 
pair a string of flux is formed compensating for the transformation 
of the quarks.  
However, the singlet expression can be interpreted as arising from
the gauge invariant Wilson loop in a particular gauge \cite{wil,bro}.
For example, evaluating a Wilson loop in the $(x_i,t)$-plane in an axial gauge,
where all links in the $i$-direction are fixed to the identity,
one obtains the chain of equations
\ba
\langle W(r,t)\rangle&=&\langle W(r,t) \delta[U_i(x),1]\rangle\nn\\
&=&\langle  L^\dag_{\alpha\beta}(\bfx)U_{i,\beta\gamma}(\bfx,\bfy)
L_{\gamma\delta}(\bfy)U^\dag_{i,\delta\alpha}(\bfx,\bfy)\delta[U_i(x),1]\rangle \nn\\
&=&
\langle  L^\dag_{\alpha\beta}(\bfx)\delta_{\beta\gamma}
L_{\gamma\delta}(\bfy)\delta_{\delta\alpha}\rangle =
\langle \tr L^\dag(\bfx)L(\bfy)\rangle ,
\ea
where $U_i(\bfx,\bfy)$ denotes the straight line of links 
in the $i$-direction connecting
$\bfx$ and $\bfy$. It represents the physical flux tube formed between the
charges, and the whole quark-antiquark-glue system is in a singlet state. 
At $T=0$, this is the only known gauge invariant operator to 
extract an interquark potential.

At finite temperature, Euclidean time is
compactified to the interval $1/T=aN_t$, 
with periodic boundary conditions for the link variables. 
The choice $t_0=1,t=N_t$ in \eq (\ref{line}) turns $L(\bfx)$ into a Polyakov loop 
encircling the torus in the time direction, transforming as
$L^g(\bfx)=g(\bfx)L(\bfx)g^{-1}(\bfx)$.
In this case, the only gauge invariant object is the 
correlation of $\tr L$, which is related to the
color averaged potential, i.e.~the weighted sum of
the singlet and adjoint channels \cite{ls,nad},
\ba \label{vav}
\ex^{-N_t V_{av}(r)}&=&\tr\left((P_1+P_{ad})C_{\bar{q}q}\right)/\tr(P_1+P_{ad})\nn\\
&=&\frac{1}{N^2}
\langle \tr L^\dag(\bfx) \tr L(\bfy)\rangle =
\frac{1}{N^2}\ex^{-N_t V_1(r)}+\frac{N^2-1}{N^2}\ex^{-N_t V_{ad}(r)}.
\ea
However, temperature can be made arbitrarily small by increasing $N_t$ at
fixed lattice spacing.
The fact that there are gauge invariant operators for the singlet and the
average potential then suggests that there should also be a gauge invariant
formulation for the adjoint channel.

The problem can be solved by dressing the source fields with gluon ``clouds'',
whose fields transform in a fundamental representation \cite{op1,op2}.
Construction of such fields is not unique. One possibility is to
use the eigenfunctions of the spatial Laplacian defined on every timeslice, 
\be \label{lev}
-\left(\Delta_i^2[U]\right)_{\alpha\beta}f^{(n)}_\beta(x)=
\lambda_n f^{(n)}_\alpha(x),  \quad \lambda^n>0.
\ee
The latter is a
hermitian operator with a strictly positive spectrum,
whose eigenvectors have the transformation property $f^{(n)g}(x)=g(x)f^{(n)}(x)$,
and may be combined into a matrix
$\Omega(x)\in SU(N)$ following the algorithm given in \cite{vw}. 
It is a non-local functional in the sense that it depends on all links
in a given timeslice.
Since \eq (\ref{lev}) only determines eigenvectors up to
a phase, this leaves a remaining freedom in $\Omega(x)$.
In the case of $SU(2)$, all eigenvalues are two-fold degenerate 
due to charge conjugation,
and the two vectors to the lowest eigenvalue are combined into $\Omega$,
which then is determined up to a global $SU(2)$ rotation $h$.
For $SU(3)$ there is no degeneracy of the eigenvalues in general. In this case
one solves for the three lowest eigenvectors to construct the  matrix $\Omega$,
which is then determined
up to a factor $h={\rm diag}(\exp(\ii \omega_1),\exp(\ii \omega_2),\exp(\ii \omega_3)),
\sum_i\omega_i=0$. 
This may be summarized by the transformation behaviour
\be \label{otrafo}
\Omega^g(x)=g(x)\Omega(x)h^\dag (t),
\ee
where $h(t)$ is free and may be different in every timeslice.

Let us then consider a dressed static quark by coupling it to such a
gluonic field, $\psi^\Omega(\bfx) \equiv \Omega^\dag(\bfx)\psi(\bfx)$. 
Replacing $\psi(\bfx)\rightarrow \psi^\Omega(\bfx)$ in \eq (\ref{pair}), one observes
the desired transformation behaviour for {\it all} $\bfx,\bfy$,
\ba 
\left (P_1\bar{\psi}^\Omega(\bfx)\psi^\Omega(\bfy)\right)^g&=&
\left(P_1\bar{\psi}^{\Omega}(\bfx)\psi^\Omega(\bfy)\right)\nn\\
\left (P_{ad}\bar{\psi}^{\Omega}(\bfx)\psi^\Omega(\bfy)\right)^g&=&
-2 \bar{T}^a\left(\psi^{\Omega \dag}(\bfx)h^\dag T^a h \psi^\Omega(\bfy)\right).
\ea
In the first line we have a singlet state. The fields $\Omega(x)$ now play the
same role as the spatial lines in the Wilson loop, allowing for flux between
the sources to compensate their transformations.
However, with this construction we have in addition a locally gauge invariant
adjoint state, which transforms as an $(N^2-1)$-plet under the residual global
symmetry of the Laplacian.
Time propagation of the dressed charge is correspondingly described by the 
dressed Wilson line
\be
\tilde{L}(\bfx)\equiv \Omega^\dag (\bfx,t_0)L(\bfx)\Omega(\bfx,t),
\la{dress}
\ee
which gauge transforms as $\tilde{L}^g(\bfx)=h(t)\tilde{L}(\bfx)h^\dag(t)$.
The correlator
$\langle \tr \tilde{L}^\dag(\bfx)\tilde{L}(\bfy)\rangle$ is gauge invariant,
and hence constitutes an alternative expression to extract the singlet potential. 
By means of the transfer matrix formalism \cite{cre,pos}
the Euclidean expectation values may be converted to traces over quantum mechanical
states in a Hilbert space formulation, elucidating the quantum mechanical 
interpretation of the correlation functions. Comparing Wilson loop and the correlator
of dressed lines, one has
\ba
\langle W(r,t)\rangle &=&
Z^{-1}\trq\{\htm^{N_t-t}\hat{\bar{\psi}}_\alpha(\bfx)\hat{U}_{\alpha\beta}(\bfx,\bfy)
\hat{\psi}_\beta(\bfy)\; \htm^t \;
\hat{\psi}^{\dag}_\gamma(\bfy)
\hat{U}^{\dag}_{\gamma\delta}(\bfx,\bfy)\}\hat{\bar{\psi}}^\dag_\delta(\bfx)\}\nn\\
\langle \tr \tilde{L}^\dag(\bfx)\tilde{L}(\bfy)\rangle &=& \nn\\ &&\hspace*{-3cm}
Z^{-1}\trq\{\htm^{N_t-t}\hat{\bar{\psi}}_\alpha(\bfx)\hat{\Omega}_{\alpha\gamma}(\bfx)
\hat{\Omega}_{\gamma\beta}^\dag(\bfy) \hat{\psi}_\beta(\bfy)
\; \htm^t \; 
\hat{\psi}^{\dag}_\delta(\bfy) \hat{\Omega}^\dag_{\delta\omega}(\bfy)
\hat{\Omega}_{\omega\tau}(\bfx)\hat{\bar{\psi}}^{\dag}_\tau(\bfx)\}.
\label{comp}
\ea
where $\hat{\psi}, \hat{\psi}^\dag$ are annihilation and 
creation operators for a static quark, $\hat{\Omega}$ acts as a multiplication 
operator and $\htm$ denotes the transfer matrix.
Inserting complete sets of eigenstates, it is evident that both 
expressions decay exponentially 
with the spectrum of the Hamiltonian, yielding identically
the same singlet potential as their ground state. 
 
Next, consider the finite temperature case, where \eq (\ref{dress}) becomes
a dressed Polyakov loop.
Since $\tr \tilde{L}(\bfx)=\tr L(\bfx)$, the expectation
value $\langle \tr \tilde{L} \rangle$ remains an order parameter for the deconfinement
transition of the pure gauge theory\footnote{
Note that 
even the untraced loop $\tilde{L}(\bfx)$ retains 
the same transformation behaviour under the center of the 
gauge group as $L(\bfx)$,
since the functionals $\Omega(x)$ are solutions of the spatial Laplacian.} 
Moreover, the average potential is identically the same as that defined from $L$. 
Substituting dressed Polyakov loops for undressed ones in \eqs (\ref{potdef}), 
we thus obtain gauge invariant, non-perturbative
definitions for the singlet, adjoint and average channels separately.  

How do the results depend on the particular
construction of the functional $\Omega[U]$?
Since $\Omega\in SU(N)$, the dressing of the source may 
also be viewed as a gauge transformation by $\Omega^\dag(x)$, and 
the corresponding operator may be written as
$\hat{\psi}^\Omega=\hat{\Omega}^\dag\hat{\psi}=
\hat{R}^{-1}(\Omega)\hat{\psi}\hat{R}(\Omega)$,
where $\hat{R}(g)$ is a unitary operator implementing a gauge transformation
by $g(\bfx)$.
The Hilbert space expression for the singlet potential then takes the form
\be
\langle \tr \tilde{L}^\dag(\bfx)\tilde{L}(\bfy)\rangle
=
Z^{-1}\trq\{\htm^{N_t-t}\hat{\bar{\psi}}_\alpha(\bfx)
\hat{\psi}_\alpha(\bfy) \hat{R}(\Omega)
\; \htm^t \;\hat{R}^{-1}(\Omega) 
\hat{\psi}^{\dag}_\beta(\bfy) 
\hat{\bar{\psi}}^{\dag}_\beta(\bfx)\}.
\ee
The above expression is identical to the undressed one 
up to a similarity transformation of the
transfer matrix, which preserves
the spectrum as well as the norm of the eigenstates. 
(Note that $\hat{R}(\Omega)$ does not commute with $\htm$ because $\Omega[U]$
depends on the link variables).
Hence, the singlet potential can in principle be extracted
by means of any $\Omega[U]$, that is local in time and transforming as 
in \eq (\ref{otrafo}).

According to the above, the dressed Wilson line 
may equivalently be viewed as a Wilson line brought to Laplacian Coulomb gauge
by a gauge transformation $g(x)=\Omega^\dag(x)$.
In the language of gauge fixing the previous statements may then be rephrased
as follows: the gauge fixed Polyakov loop correlation function in the singlet 
channel falls off with gauge invariant 
eigenvalues of the Hamiltonian. These may be extracted in 
any unique gauge that is local in time, i.e.
preserves the spectrum of the transfer matrix. This is true, for example, for the
Coulomb gauge, but not for the Landau or complete Laplacian gauge, 
which depend on all time-like links as well. 

In the remainder numerical results will be presented supporting these
theoretical statements and illustrating their numerical feasibility.
This first exploratory study is done for SU(2) 
in 2+1 dimensions, for its significantly lower numerical cost and 
fast continuum approach. 
In this case the coupling constant $g^2$ has dimension of mass and provides a
scale. As a consequence, the short distance
Coulomb part of the potential defined through the Wilson loop 
is logarithmic, and in next-to-leading order
a contribution to linear confinement is obtained even in perturbation 
theory,
\be
V^{3d}_{pert}(r)=\frac{g^2C_F}{2\pi}\ln(g^2r)+\sigma_{pert}r+\op(g^6r^2),
\ee
where $\sigma_{pert}$ however falls short of the non-perturbative 
string tension by a factor 
of about 2/3 \cite{york}. Since the group theory 
is unaffected by the number of space dimensions, the leading order 
perturbative relation between singlet and adjoint Coulomb terms, \eq (\ref{pert}),
carries over to this case (where the SU(2) adjoint channel corresponds to a triplet,
$V_{ad}=V_3$). 
The lattice gauge coupling is given by $\beta=4/(ag^2)$.

First, it is instructive to test the numerical feasibility of the 
dressed Wilson line by comparing its zero temperature correlation function
with the Wilson loop, as in \eq (\ref{comp}). \fig \ref{wl} shows the 
singlet potential extracted from those
correlation functions, and numerically confirms their analytically established
equality. (Numerically this has also been observed for Wilson loop and lines
in the adjoint representation \cite{fp}).

\begin{figure}[ht]
\vspace*{0.7cm}
\centerline{\epsfxsize=8cm\hspace*{0cm}\epsfbox{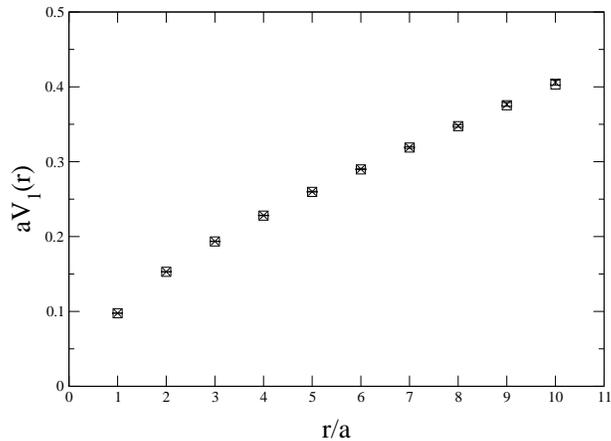}}

\caption[a]{{\em
Singlet potential extracted from the Wilson loop $\langle W(r,t) \rangle$ 
(open squares), and correlations of dressed Wilson lines 
$\langle\tilde{L}(\bfx)\tilde{L}(\bfy)\rangle$ (crosses), 
for $\beta=9, L^2\cdot N_t=32^3$. 
}}
\la{wl}
\end{figure}

Next, we consider finite temperatures and dressed Polyakov loop correlations.
The deconfinement transition of the 2+1 dimensional gauge theory has been studied in 
\cite{Teper:1993gp},\cite{Engels:1997dz}. The relationship between the 
inverse critical temperature
in lattice units, $N_t^c=1/(aT_c)$, and the lattice gauge coupling $\beta$ is well 
fitted by $N_t^c(\beta)=(\beta-0.37)/1.55$ \cite{Hart:2000en}.
For the following calculation of the static potentials we work at $\beta=9$, for which
the deconfinement transition is between $N_t^c=5,6$. 
The three potentials are shown for decreasing temperatures
in \fig \ref{pots}.

\begin{figure}[th]

\centerline{\epsfxsize=6cm\hspace*{0cm}\epsfbox{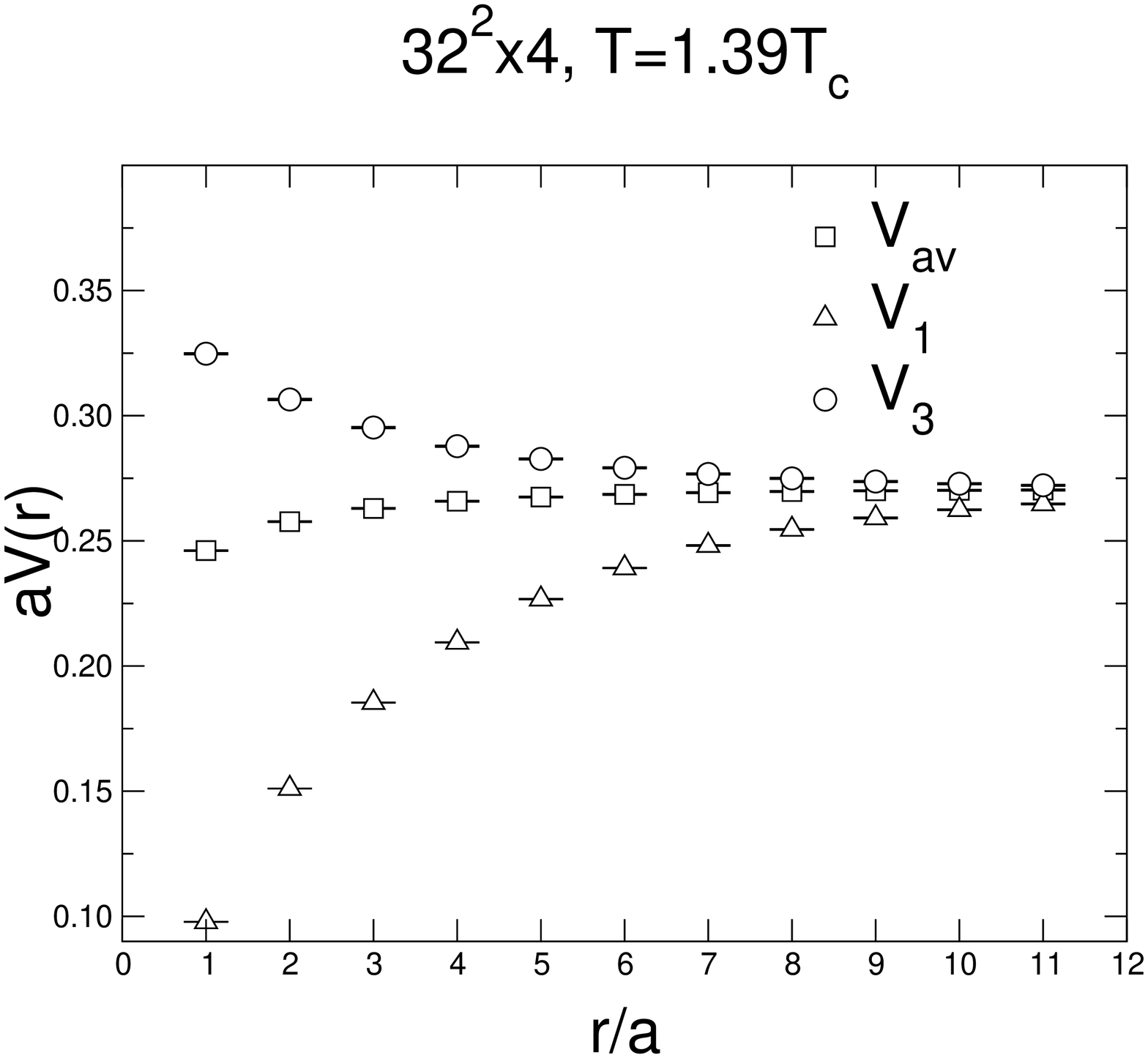}%
\epsfxsize=6cm\hspace*{1cm}\epsfbox{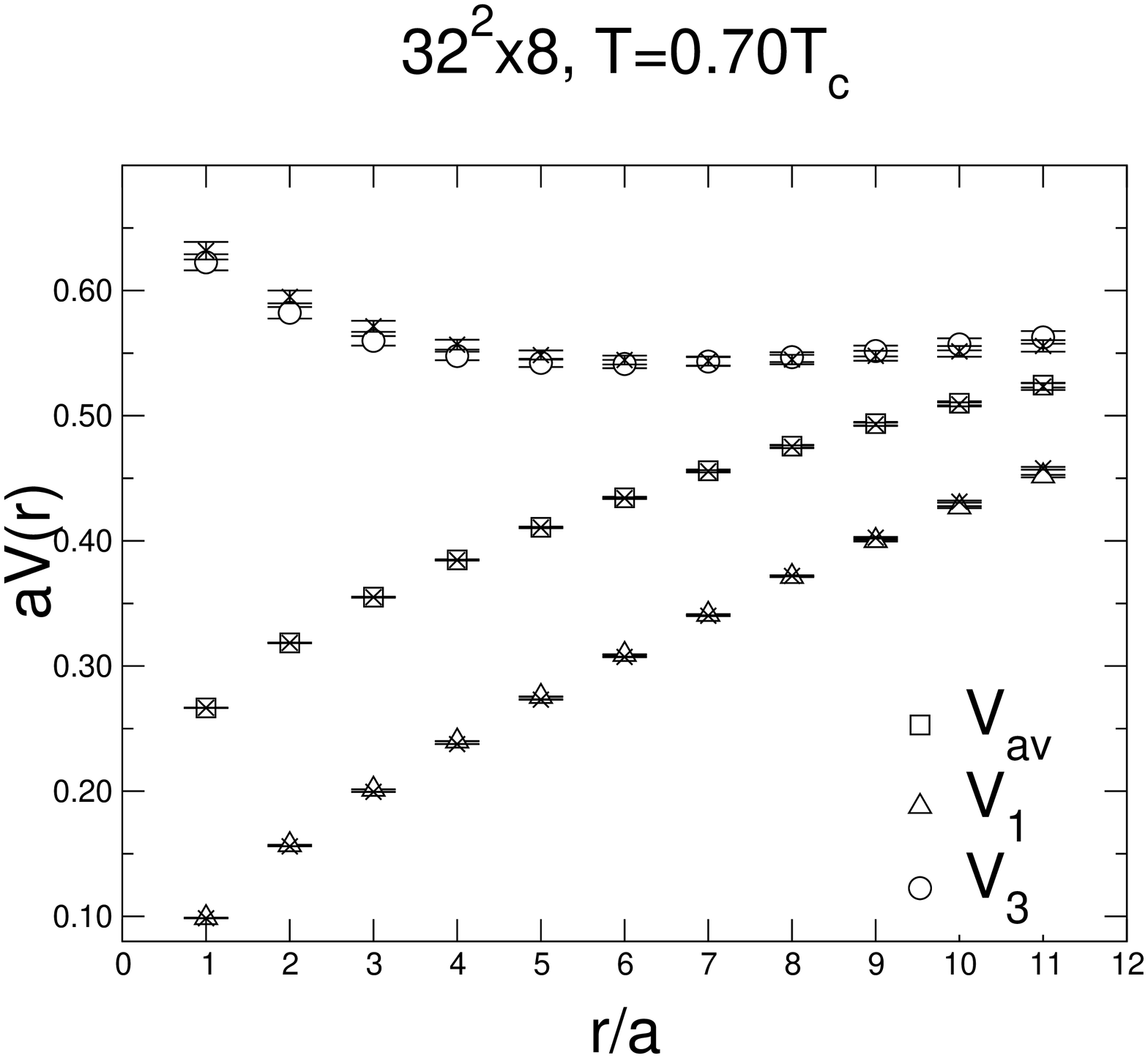}}
\vspace*{0.5cm}
\centerline{\epsfxsize=6cm\hspace*{0cm}\epsfbox{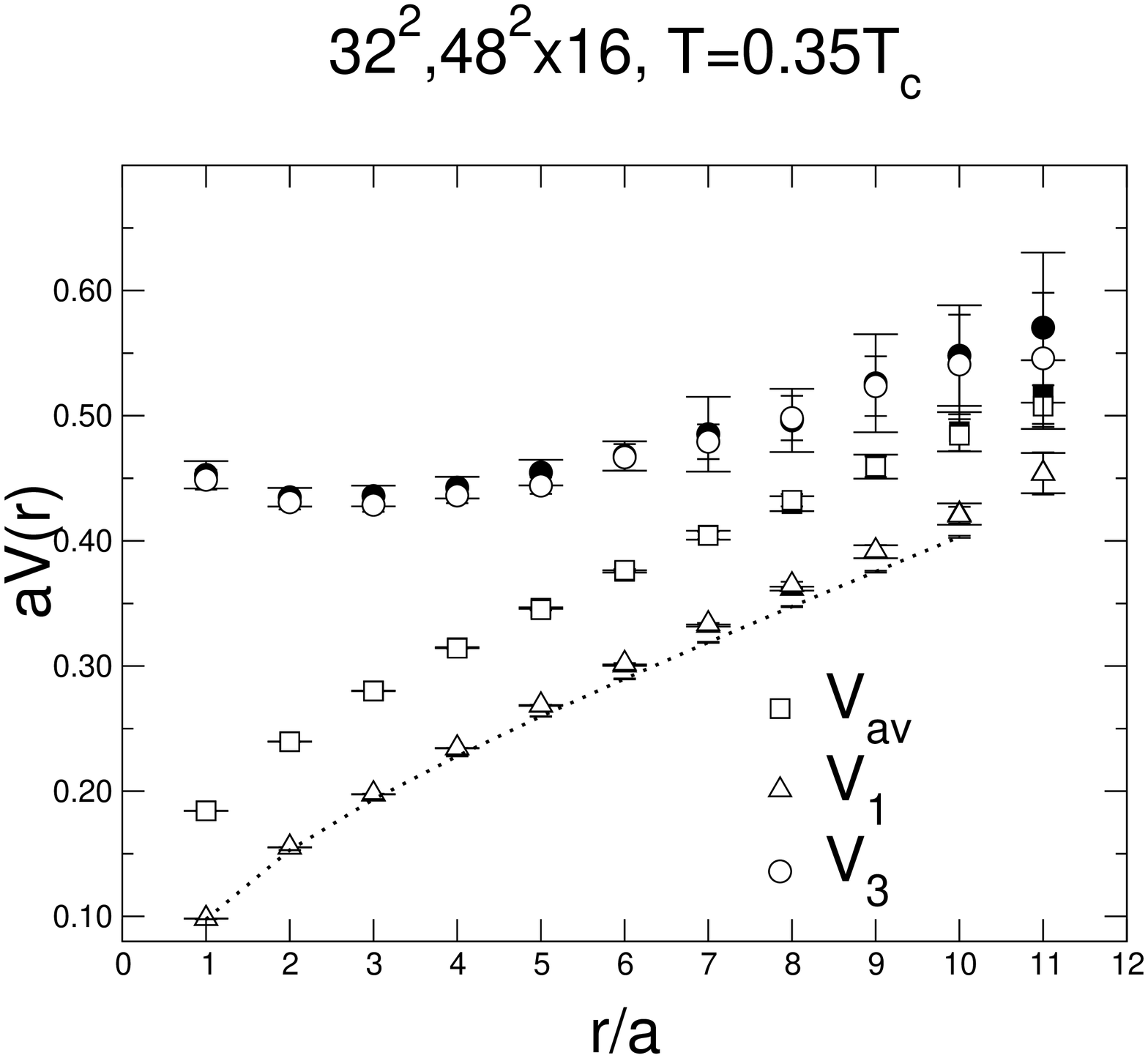}}

\caption[a]{{\em Singlet, triplet and average potentials calculated
for different temperatures at $\beta=9$. At $T=0.7T_c$, the crosses denote
data obtained in standard Coulomb gauge. At $T=0.35T_c$, filled symbols
denote data obtained on a $48^2\cdot 16$ lattice. The dotted line marks
the $T=0$ Wilson loop data from \fig \ref{wl}.  
}}
\la{pots}
\end{figure}

The results clearly confirm the perturbative expectation
that the triplet potential is repulsive at short 
distances. 
In the deconfined phase all potentials assume a Yukawa 
form and merge at a common constant value
at large distance, i.e. all color interactions are screened.
In the confinement phase, however, the triplet potential becomes attractive at 
larger distances, while staying
always above the singlet potential for the 
distances covered in the simulation. 
This is in accord with the 
fact that no stable charged states are observed empirically. 
Nevertheless, at low temperatures a potential well seems to form
in which a metastable bound state appears possible.
An investigation of this phenomenon in 4d SU(3) would be
particularly interesting for heavy quark phenomenology, such as
$J/\Psi$-production through an octet channel.
A nice consistency check for the calculation is provided by
comparing the singlet potential in the low temperature case with the 
zero temperature potential obtained from the Wilson loop.
The latter is indicated in the figure for $T=0.35T_c$ by the dotted line,
which is approached by the low $T$ Polyakov loop data from above, as expected
on physical grounds.

As reported in \cite{op2}, 
the non-locality of the functionals $\Omega[U]$ has them project predominantly on
flux tubes encircling the periodic spatial volume, hampering the observation of
localized states by large finite volume effects. 
Such effects are absent in the current
calculations, as is demonstrated in \fig \ref{pots} in the low $T$ plot,
where the data for $L=32,48$ yield entirely consistent results.
This is easily understood recalling that the quark-antiquark system is
spatially extended, and the non-locality of the $\Omega(x)$ actually enables
flux tubes to connect the pair over large distances. 
The flux represented by $\Omega(x)$ also closes through the periodic boundary
conditions. As a consequence, we really measure a superposition of 
$\exp -N_tV(r)$ and $\exp -N_tV(L-r)$. However, for our choices of $N_t,L$, the
second term is exponentially suppressed compared to the first one, and no finite
volume effects are observed.

In order to illustrate the independence
of the calculation of the particular choice of $\Omega$, 
the $T=0.7T_c$  plot in \fig \ref{pots}
shows an additional calculation, where $\Omega[U]$ is the 
functional to fix 
standard Coulomb gauge, i.e. it minimizes the 
function $R[U]=\sum_{x,i}[1-1/N\tr \Omega^\dag(x)U_i(x)\Omega(x+\hat{\i})]$ 
in every timeslice.
The corresponding data points, shown as crosses in the plot, 
coincide with those from the previous calculation.
The reader is reminded that the lattice Coulomb gauge in principle is afflicted
by Gribov copies.
Since we know from our analytical considerations that the 
potentials have to be independent of the
choice of $\Omega$, this numerical result may be 
interpreted as a test for the effect of Gribov copies
in the standard Coulomb gauge.

Having separate non-perturbative results for the singlet and adjoint
channel potentials, they should approach the leading order perturbative
prediction, \eq (\ref{pert}), in a regime where
perturbation theory is valid. This is the case at short distances, and
expected to extend to larger distances 
in the deconfined phase. Since the static potential
includes an unphysical divergence which is different in continuum
and lattice regularizations, we compare the physical forces
$F_i=dV_i/dr$. \fig \ref{pt} shows results obtained in the deconfined phase
on a finer lattice, by employing the discretized
forward derivative $F_i(r)=\Delta_rV_i(r)=(V_i(r+a)-V(r))/a$.
Satisfactory agreement with the leading order perturbative result
is observed.
 
\begin{figure}[th]
\vspace*{0.7cm}
\centerline{\epsfxsize=8cm\hspace*{0cm}\epsfbox{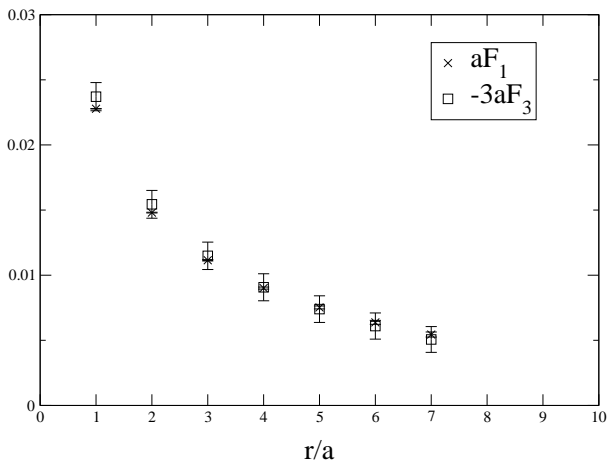}}

\caption[a]{{\em
Comparison of the forces in the singlet and adjoint channels, computed at
$\beta=18, L^2\cdot N_t=48^2\cdot 8$, corresponding to $T=1.42T_c$. 
}}
\la{pt}
\end{figure}

To conclude, in following the methods of \cite{op1,op2}, we have shown that
a dressed static quark-antiquark pair can exist in locally gauge invariant singlet
{\it and} adjoint states, the latter
transforming as a global $(N^2-1)$-plet.
Calculation of the correspondingly dressed Polyakov loop correlation functions
then permits the non-perturbative extraction of the 
singlet, adjoint and average potentials which are each 
eigenvalues of the Hamiltonian and hence gauge invariant.
Numerical results for 2+1 d SU(2) gauge theory provide a consistent picture,
in agreement with perturbation theory at small distances and large temperatures.
Moreover, the calculation allows to observe how the finite temperature 
singlet potential extracted from the dressed Polyakov loop
approaches the Wilson loop results in the zero temperature limit.
The adjoint channel potential is repulsive at short distances. At zero
temperature it becomes confining at large distances, allowing for a metastable bound
state which should be of interest for heavy quark phenomenology.
The considerations in this work can be applied to construct gauge invariant
symmetric and antisymmetric quark-quark potentials as well.
Apart from their relevance for heavy quark physics and the confinement problem, 
our results should also offer
new ways to study the forces and excitations in hot and dense QCD.
\\

\noindent
{\bf Acknowledgements:}
I enjoyed useful discussions with O.~B\"ar, Y.~Schr\"oder, 
B.~Svetitsky and U.-J.~Wiese.
The simulations in this paper were performed on a NEC/SX-32 at the HLRS at the
Universit\"at Stuttgart, supported by the ITP, Universit\"at Heidelberg.


\begin{thebibliography}{99}

\bibitem{bro}
L.~S.~Brown and W.~I.~Weisberger,
Phys.\ Rev.\ D {\bf 20} (1979) 3239.

\bibitem{nad}
S.~Nadkarni,
Phys.\ Rev.\ D {\bf 34} (1986) 3904.

\bibitem{att}
N.~Attig et al., 
Phys.\ Lett.\ B {\bf 209} (1988) 65.

\bibitem{som}
S.~Necco and R.~Sommer,
Phys.\ Lett.\ B {\bf 523} (2001) 135.

\bibitem{ls}
L.~D.~McLerran and B.~Svetitsky,
Phys.\ Rev.\ D {\bf 24} (1981) 450.

\bibitem{op1}
O.~Philipsen,
Phys.\ Lett.\ B {\bf 521} (2001) 273.

\bibitem{op2}
O.~Philipsen,
hep-lat/0112047, to appear in Nucl.\ Phys.\ B.

\bibitem{po}
A.M. Polyakov (ICTP, Trieste). IC-78-4-mc (microfiche), 64 pp., Feb 1978,
Lectures given at ICTP, Trieste, Nov. 1977;\\
G.~'t Hooft,
Nucl.\ Phys.\ B {\bf 153} (1979) 141.

\bibitem{wil}
K.~G.~Wilson,
Phys.\ Rev.\ D {\bf 10} (1974) 2445.

\bibitem{vw}
J.~C.~Vink and U.~Wiese,
Phys.\ Lett.\ B {\bf 289} (1992) 122.

\bibitem{cre}
M.~Creutz,
Phys.\ Rev.\ D {\bf 15} (1977) 1128.

\bibitem{pos}
M.~L\"uscher,
Commun.\ Math.\ Phys.\ {\bf 54} (1977) 283.

\bibitem{york}
Y.~Schr\"oder, {\it The static potential in $QCD_3$ at one loop}, 
proceedings of `Strong and Electroweak Matter 97',  Eger, Hungary,
World Scientific 1998, p.394.

\bibitem{fp}
P.~de Forcrand and O.~Philipsen,
Phys.\ Lett.\ B {\bf 475} (2000) 280.

\bibitem{Teper:1993gp}
M.~Teper,
Phys.\ Lett.\ B {\bf 313} (1993) 417.

\bibitem{Engels:1997dz}
J.~Engels, F.~Karsch, E.~Laermann, C.~Legeland, M.~L\"utgemeier, B.~Petersson and T.~Scheideler,
Nucl.\ Phys.\ Proc.\ Suppl.\  {\bf 53} (1997) 420.

\bibitem{Hart:2000en}
A.~Hart, B.~Lucini, Z.~Schram and M.~Teper,
JHEP {\bf 0006} (2000) 040.

\end{thebibliography}
\end{document}